\documentclass[12pt, draftclsnofoot, onecolumn]{IEEEtran}

\ifCLASSINFOpdf
\else
\fi

\usepackage{graphics,graphicx} 
\usepackage{epsfig} 
\usepackage[final]{pdfpages}

\usepackage{amsmath}
\usepackage{amssymb}
\usepackage{tikz}
\usepackage{amsthm}
\usepackage{comment}
\usepackage[export]{adjustbox}
\usepackage{mathcomSTEP}
\usepackage{subcaption,setspace}
\usepackage[utf8]{inputenc}

\usepackage{tikz}
\usetikzlibrary{positioning}
\usetikzlibrary{decorations.pathreplacing,angles,quotes}

\allowdisplaybreaks

\newtheorem{thm}{Theorem}[section]

\newtheorem{lem}[thm]{Lemma}
\theoremstyle{remark}
\newtheorem{rem}[thm]{Remark}

\newtheorem{definition}{Definition}

\usetikzlibrary{positioning}
\usetikzlibrary{arrows,fit}
\usetikzlibrary{decorations.pathreplacing}
\tikzstyle{int}=[draw, fill=blue!10, minimum height = 1cm, minimum width=1.5cm,thick ]
\tikzstyle{rbox}=[draw, rounded corners=5 pt, minimum height = 1cm, minimum width=1.5cm,thick ]
\tikzstyle{sum}=[circle, fill=blue!10, draw=black,line width=.5 pt,minimum size = 0.05 cm, thin ]
\tikzstyle{joint} = [draw, circle, minimum size=1em]
\usetikzlibrary{shapes,arrows,backgrounds,positioning}
\usetikzlibrary{positioning,intersections,decorations.pathreplacing}
\tikzstyle{int}=[draw, fill=blue!10, minimum height = 1.2cm, minimum width=1.5cm,thick ]
\tikzstyle{cw}= [fill=blue!10, draw, rounded corners = 1 ex,minimum height = 1cm]
\tikzstyle{joint} = [draw, circle, minimum size=1em]

\tikzset{
  treenode/.style = {align=center, inner sep=0pt, text centered,
    font=\sffamily},
    arn_n/.style = {treenode, circle, draw=black, text width=2 em}
    arn_nn/.style = {treenode, circle, draw=black, text width=2.5 em}
}

\begin{document}

\title{
On Capacity of  \\
the Writing onto Fast Fading Dirt Channel
}

\author{%
\IEEEauthorblockN{%
Stefano Rini
and Shlomo Shamai (Shitz)
}
%
%

%
\thanks{
The work of S. Rini was funded by the  Ministry Of Science and Technology (MOST) under the grant 103-2218-E-009-014-MY2.
The work of S. Shamai was supported by the Israel Science Foundation (ISF) and by the European FP7 NEWCOM\#.
}
}

\maketitle

\begin{abstract}
The ``Writing onto Fast Fading Dirt'' (WFFD) channel is investigated to study the effects of partial channel knowledge on the capacity of the ``writing on dirty paper'' channel.
The WFFD channel is  the Gel'fand-Pinsker channel in which the output is obtained as the sum of the input, white Gaussian noise and a fading-times-state term.
The fading-times-state term is equal to the element-wise product of the channel state sequence, known only at the transmitter, and a fast fading process, known only at the receiver.
We consider the case of  Gaussian distributed channel states and derive an approximate characterization of  capacity for
different classes of fading distributions, both continuous and discrete.
In particular, we prove that if the fading distribution concentrates in a sufficiently small interval, then capacity is approximately equal to  the  AWGN capacity times the probability of this interval.
We also show that there exists a class of  fading distributions for which having the transmitter treat the fading-times-state term as additional noise closely approaches capacity.
Although a closed-form expression of the capacity of the general WFFD channel remains unknown,
 our results show that the presence of fading can severely reduce the usefulness of  channel state knowledge at the transmitter.
\end{abstract}


\section*{Introduction}
\label{sec:Introduction}
The classic ``Writing on Dirty Paper'' (WDP) channel capacity result \cite{costa1983writing}  establishes that full state pre-cancellation can be attained in the Gel'fand-Pinsker (GP) channel with additive state and additive white Gaussian noise, regardless of the  distribution of the state sequence.
Albeit very promising, this result assumes that perfect channel knowledge is available at the users:
this assumption  does not hold in many communication scenarios in which channel conditions vary over time and with limited feedback between the receiver and the transmitter.
For this reason, we wish to investigate the effects of partial channel knowledge on the performance of state pre-cancellation.
More specifically, we study the capacity of the ``Writing onto Fast Fading Dirt'' (WFFD) channel, a variation of Costa's WDP channel in which the state sequence is  multiplied by a fast fading process known only at the receiver.

The WFFD channel models the  downlink transmission scenario in which a base station wishes to communicate to a receiver in the presence of an interferer.
The base station acquires the  message sent by the interferer through the network architecture while the receiver learns the channel toward  the interferer from the pilot tones broadcasted by the interferer.
Due to rate limitations in the control and feedback channels, the transmitter and the receiver are unable to exchange each other's knowledge.
This, therefore, results in the situation in which the transmitter knows the interfering message but not the interfering channel, while the receiver knows the interfering channel but not the interfering message.
For this scenario, one wishes to determine the limiting interference cancellation performance that are attainable despite the partial and asymmetric system knowledge at the transmitter and the receiver.
\subsubsection*{Related Results}
The GP channel \cite{GelFandPinskerClassic} is the point-to-point channel in which the  output is obtained as a random function of
the input and a state sequence which is provided non-causally at the transmitter.
The capacity of this model  is  expressed in \cite{GelFandPinskerClassic} as the maximization of a non-convex function for which the optimal solution is not easily determined, either explicitly or through numerical evaluations.
For this reason, very few closed-form expressions of the GP channel capacity are available  in the literature.
One of the few models for which capacity is known in closed-form is the WDP channel:
in \cite{costa1983writing} Costa
shows, perhaps surprisingly, that the capacity of the WDP channel is equal to the capacity of the Gaussian point-to-point channel.
This result implies that it is possible for the encoder  to fully pre-code its transmissions against the known channel state.
In the literature, few authors have investigated extensions of the result in \cite{costa1983writing} to include fading and partial channel knowledge.

The ``Carbon Copying onto Dirty Paper'' (CCDP) channel \cite{LapidothCarbonCopying} is the $M$-user compound channel in which the output at each compound receiver is obtained as the sum of the input, Gaussian noise and one of $M$ possible state sequences, all non-causally known only at the transmitter.
When the state sequences at each receiver are scaled versions of the same sequence, the CCDP channel models the
WDP channel in which the channel state is multiplied by a slow fading process.

The WDP channel in which both the input and the state sequence are multiplied by the same fading realization is studied in  \cite{zhang2007writing}.
The authors consider both the case of fast and slow fading and evaluate the achievable rates using Costa pre-coding,
showing that the rate loss from full state pre-cancellation is  vanishing in both scenarios as the transmit power grows to infinity.
The model in which the same fading realization multiplies both the  input and the  state is closely related to the fading broadcast channel as, in the latter model,  the channel state models the codeword intended for another user in the network:
in \cite{bennatan2008fading} and \cite{hindy2016lattice}, lattice coding strategies  are employed to derive an achievable region for this model.

The WDP channel in which slow fading affects only  the state sequence is first studied in \cite{grover2007need} for the case of phase fading.
In \cite{grover2007need}, the first inner and outer bounds to capacity are obtained while \cite{grover2007writing} studies the outage probability for this model.
In \cite{rini2014impact}, we show the approximate capacity of this model for some classes of the phase fading.
Achievable rates under Gaussian signaling are derived in  \cite{avner2010dirty} for the case of Gaussian distributed  fast fading.
These attainable rates are also compared to lattice coding strategies and some numerical observations are provided on the performance of these  various coding choices.
The performance of lattice coding strategies for this channel model is further studied  in \cite{bennatan2009bounds},
\cite{khina2010robustness}  and \cite{albounds}.
In \cite{bennatan2009bounds},  the authors also derive an  upper  bound  on  the maximum achievable transmission rates which is shown
to be tight in some settings.
In  \cite{khina2010robustness}, it is shown that randomizing the state scaling at the transmitter can improve the performance over a pre-determined scaling choice.
%
%
%
\subsubsection*{Contributions}
\label{sec:Contributions}
We investigate the capacity of the WFFD channel\footnote{In the literature this channel has also been  referred to as  ``dirty paper channel with fading dirt'', ``writing on  fast faded dirt'' and  ``dirty paper coding channel with fast fading''. We prefer the term ``writing on fast fading dirt'' for both brevity and clarity.}
in which the state sequence is a white Gaussian sequence:
we consider, separately, the case of discrete and continuous fading distributions.
A summary of the contributions for these two scenarios is as follows:

\smallskip
\noindent
$\bullet$ {\bf Sec. \ref{sec:Discrete fading distribution}-- Discrete fading distributions.}
  We begin by determining capacity to within a constant gap for the case of uniform antipodal fading. For this simple fading distribution capacity can be approached   by transmitting the superposition of two codewords: the bottom codeword treats the fading-times-state as additional noise while the top codeword is pre-coded against one of the fading realizations  times the channel state.
This result is extended to two classes of fading distributions: the class of distributions with  mode larger than a half and
 the class of uniform distributions with exponentially spaced points in the support.
 In both cases, capacity is approached to within a small gap by a combination of superposition coding and state pre-coding as in the case of uniform antipodal fading.

\smallskip
\noindent
$\bullet$ {\bf Sec. \ref{sec:WFFD with continuos fading}-- Continuous fading distributions.}
We begin by considering the case of a symmetric continuous fading distribution and  show simple conditions under which capacity is at most half of the AWGN capacity.
We then derive the approximate capacity for the case of a continuous fading distribution which concentrates around a sufficiently narrow interval.
The converse proof is shown by  relating the capacity of the model
with continuous fading to the capacity of the model in which the fading distribution is a quantized version of the original distribution.
Finally, we show that there exists a heavy-tailed fading distribution for which the capacity of the WFFD channel is approximatively equal to the capacity of the channel without state knowledge.
%

The main theoretical contributions of the paper consist in the development of new outer bounding techniques to characterize the capacity of a model comprising both channel states and partial channel knowledge.
On the other hand, the inner bounds used throughout the paper are rather straightforward combinations of Costa pre-coding and superposition coding.
From a high level perspective, this shows that in the instances we consider
state pre-coding is substantially rendered ineffective by the presence of channel uncertainty.
Although this conclusion  does not hold in general,  our results partially reveal the conditions under which robust state pre-cancellation is no longer possible.

\subsubsection*{Paper Organization}
\label{sec:Paper Organization}
The remainder of the paper is organized as follows:
in Sec. \ref{sec:Channel Model} we introduce the channel model under consideration.
Sec. \ref{sec:Related Results} presents relevant results available in the literature.
Sec. \ref{sec:Discrete fading distribution} considers the case of discrete fading distributions while  Sec. \ref{sec:WFFD with continuos fading} studies the case of continuous distributions.
Finally, Sec. \ref{sec:Conclusion} concludes the paper.
%
\section{Channel Model}
\label{sec:Channel Model}
\begin{figure}
\centering
\begin{tikzpicture}[node distance=2.5cm,auto,>=latex]
  \node at (0,0) (source) {$W$};
  \node [int] (enc) [right of = source, node distance = 1.5 cm]{Enc.};
   \node (Pyx) [joint, right of = enc, node distance = 2 cm]{};
   \node (Pyx) [right of = enc, node distance = 2 cm]{+};
   \node (Pyx2) [joint, right of = Pyx, node distance = 1.5 cm]{};
   \node (Pyx2) [right of = Pyx, node distance = 1.5 cm]{+};
    \node [int] (dec) [right of = Pyx2, node distance = 2    cm]{Dec.};
    \node  (dest) [right of=dec, node distance = 1.5 cm] {$\Wh$};
\node (mul) [joint, below of = Pyx, node distance = 1 cm]{};
\node (mul) [below of = Pyx, node distance = 1 cm]{$\circ$};
\node (noise) [above of = Pyx2, node distance = 1.5 cm]{$Z^N$};
\node (a) [below of = mul,  node distance = 1.25 cm]{  };
\node (state) [left of = a,  node distance =.75 cm]{$S^N$};
\node (fading) [right of = a,  node distance =.75 cm]{$A^N$};

   \draw[->,line width=1.5 pt] (source) -- (enc);
   \draw[->,line width=1.5 pt] (dec) -- (dest);
   \draw[->,line width=1.5pt] (enc) -- node[above] {$X^N$}(Pyx);
   \draw[->,line width=1.5pt] (Pyx2) -- node[above] {$Y^N$}(dec);
   \draw[->,line width=1.5pt] (Pyx) -- (Pyx2);
   \draw[->,line width=1.5pt] (mul) -- (Pyx);
   \draw[->,line width=1.5pt] (state) --  (mul);
   \draw[->,line width=1.5pt] (fading) --  (mul);
   \draw[->,line width=1.5pt] (noise) -- (Pyx2);
   \draw[->,line width=.75pt,dashed] (state) -|(enc);
\draw[->,line width=1.5pt] (fading) -| (dec);
%
 \end{tikzpicture}
\caption{The ``Writing onto Fast Fading Dirt'' (WFFD) channel.
}
\label{fig:WFFDgeneral}
\vspace{-.5 cm}
\end{figure}
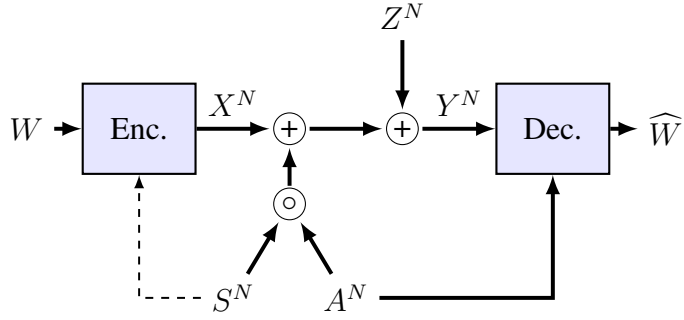

The WFFD channel, also depicted in Fig. \ref{fig:WFFDgeneral}, is the GP channel in which the output is obtained as
\ea{
Y^N=X^N+c A^N \circ S^N+Z^N,
\label{eq:Dirty Paper Channel}
}
 where $X^N$ denotes the channel input, $S^N$ the channel state, $A^N$ the fading sequence and $Z^N$  the additive noise while $\circ$ indicates the Hadamard, or element-wise, product\footnote{In other words, $A^N \circ S^N=[ A_1 S_1, A_2 S_2 \ldots A_N S_N ]^T$.}.
Having knowledge of the channel state $S^N$, the encoder wishes to reliably communicate the message $W\in \Wcal= [1 \ldots 2^{NR}]$ to
 the receiver through the channel input $X^N$.
Upon receiving the channel output $Y^N$ and the fading realization $A^N$, the receiver produces the estimate $\Wh \in \Wcal$ of the
transmitted message\footnote{Note that fading sequence $A^N$ can be seen as an additional channel output at the receiver,
together with $Y^N$ in \eqref{eq:Dirty Paper Channel}.}.
The channel input $X^N$ is subject to the second moment constraint
$\Ebb \lsb |X_i|^2 \rsb  \leq P, \ \forall \  i \ \in [1\ldots N]$.
Both the channel state and the additive noise are white Gaussian sequences, i.e. $Z^N,S^N \sim  i.i.d. \ \Ncal(0,1)$ while the %
fading sequence $A^N$ is an i.i.d. sequence from the distribution $P_A(a)$, with support $\Acal$, either continuous or discrete.
%
Without loss of generality we further assume $\var[A]=1$ and $c \in \Rbb^+$.

In the study of the WFFD channel, standard definitions of  code, achievable rate and capacity are employed \cite{ThomasCoverBook}.
\begin{definition}{\bf Code, probability of decoding error, achievable rate.}
\label{def:code}
A $(2^{N R};N)$ code for the  WFFD channel consists of
an encoding  and a decoding function, $X^N = f(W, S^N)$ and  $\Wh = g(Y^N,A^N)$ respectively.
The probability of error for a  $(2^{N R}; N)$ code, $P_e(2^{N R}; N)$, is defined as
\ea{
P_e(2^{N R}; N)
=\Pr \lsb  \Wh (Y^N, A^N)\neq W \rsb,
\label{eq:pe}
}
%
where the probability in the RHS of \eqref{eq:pe} is also averaged over all fading  and state  sequences and transmitted messages.
A rate $R\in \Rbb^+$ is said to be  achievable if there exists a sequence of codes such that the probability of error $P_e(2^{N R}; N)$ goes to zero as  $N$ goes to infinity.
\end{definition}

\begin{definition}{\bf Capacity and  approximate capacity.}
The capacity $\Ccal$ is the supremum of all the achievable rates.
%
An inner bound $R^{\rm IN}$ and an outer bound $R^{\rm OUT}$ to capacity for which
\ea{
R^{\rm OUT}-R^{\rm IN} \leq \Delta,
}
for some constant $\Delta \in \Rbb^+$, are said to characterize the capacity to within an additive gap of $\Delta$ bits--per--channel--use ($\bpcu$) or,
for brevity, to determine the approximate capacity to within $\Delta \ \bpcu$.
\end{definition}
Since the WFFD channel is a special case of the GP channel,  its capacity is obtained as
\ea{
\Ccal=\max_{P_{U,X|S}} \ I(Y; U|A) - I(U;S).
\label{eq:Capacity of GP channel}
}
The expression in \eqref{eq:Capacity of GP channel} is convex in $P_{X|S,U}$ for a fixed $P_{U| S}$  which implies that $X$  can be chosen to be a deterministic function of $U$ and $S$.
On the other hand, \eqref{eq:Capacity of GP channel} is neither convex nor concave in $P_{U|S}$ for a fixed $P_{X|S, U}$: accordingly, determining a closed-form solution for the maximization in \eqref{eq:Capacity of GP channel} is generally challenging.
Additionally, the lack of tight bounds on the cardinality of the auxiliary random variable  $U$ further complicates the task of obtaining numerical approximations of the optimal solution.
For these reasons, in the following we provide alternative inner and outer bounds to capacity which are expressed only as a function of $P,c$ and $P_A(a)$.
We also  determine the approximate capacity  for some class of distributions, focusing on those instances in which a simple combination of known achievable strategies is sufficient to achieve capacity.

In the remainder of the paper, we refer to the term $c A^N \circ S^N$ as the ``fading-times-state'' term and  use  the parameter $c$ to normalize the variance
of both the state and the fading distributions to one.
\begin{lem}{\bf Mean and variance of the fading-time-state term.}
\label{lem:Mean and variance of the fading-time-state term}
For the model in \eqref{eq:Dirty Paper Channel}, the variance of the state and the fading distributions are taken unitary without loss of generality. Also,  the channel state is taken to have zero mean and $c \in \Rbb^+$ without loss of generality.
\end{lem}
\begin{IEEEproof}
The proof is omitted for brevity.
\end{IEEEproof}

The following lemma is useful in the tightening of certain outer bounds derived in the following.

\begin{lem}
\label{lem:capacity decreasing c}
The capacity of the WFFD channel is decreasing in the parameter $c$.
\end{lem}
\begin{IEEEproof}
The proof is  presented in App. \ref{app:capacity decreasing c}.
\end{IEEEproof}
\section{Related Results}
\label{sec:Related Results}

This section briefly introduces some  results available in the literature which  are relevant to the study of the WFFD channel.

\medskip
\noindent
$\bullet$ {\bf  The ``Writing on Dirty Paper'' (WDP) channel.}
One of the few GP channel models for which the maximization in \eqref{eq:Capacity of GP channel} is known in closed-form is the WDP channel.
For this model, the optimal assignment in \eqref{eq:Capacity of GP channel} is
\ea{
&
X \sim \Ncal(0,P), \ \  X \perp S
\nonumber\\
& U= X + \f {P} {P+1}S,
\label{eq:DPC assigment}
}
and yields $\Ccal=1/2 \log(1+P)$, regardless of the distribution of $S^N$.
The assignment in \eqref{eq:DPC assigment} is usually referred to as  ``Dirty Paper Coding'' (DPC).

\medskip
\noindent
$\bullet$ {\bf The ``Carbon Copying onto Dirty Paper'' (CCDP) channel.}
The CCDP channel \cite{LapidothCarbonCopying} is the $M$-user compound channel in which a  channel output is obtained as the sum of the input, Gaussian noise and one of $M$ Gaussian state sequences.
The transmitter has non-causal knowledge of all of the $M$ state sequences while the receivers have no additional knowledge.
%
More specifically,
\ea{
Y_m^N = X^N +  c S_m^N + Z_m^N, \quad m \in [1 \ldots M],
\label{eq:Carbon copying}
}
where $S_m^N \sim \iid \Ncal(0,Q_m)$ and $\{ S_m^N, \ m \in [1\ldots M]\}$ have any jointly Gaussian distribution.
In \cite{LapidothCarbonCopying}, the authors derive the first inner and outer bound for this model.
The approximate capacity for the case of $M=2$ and independent, unitary variance \footnote{As in Lem. \ref{lem:Mean and variance of the fading-time-state term}, the assumption of unitary variance is without loss of generality.} is derived as \cite{rini2014strongFading}.
\begin{thm} {\bf Outer bound and approximate capacity for the 2-user CCDP channel with independent  states \cite{rini2014strongFading}.}
\label{th:Outer bound for the two-user, independent state case }
The capacity of the 2-user CCDP channel with $S_1^N,S_2^N \sim \iid \ \Ncal(0,1)$, $S_1^N \perp S_2^N$  is upper bounded as
\ea{
\Ccal \leq R^{\rm OUT} =
\lcb \p{
\f 12 \log \lb 1 + P \rb +1/2  & c^2 \leq 2 \\
\f 12 \log \lb \f{P+c^2/2+1}{c^2}\rb & \\
\ \ \  + \f 14 \log \lb \f {c^2} 2 \rb+1/2 & 2 \leq c^2 <2(P+1) \\
\f 14 \log (P+1) & c^2 \geq 2(P+1),
}
\rnone
\label{eq:outer lapidoth}
}
and the  capacity is to within $1 \ \bpcu$ from the outer bound in \eqref{eq:outer lapidoth}.
\end{thm}
Capacity in Th. \ref{th:Outer bound for the two-user, independent state case } is approached by sending the superposition of two codewords:
the base codeword treats the states as additional noise  while the top codeword is pre-coded against each of the state realizations for half of the time.
Th. \ref{th:Outer bound for the two-user, independent state case } shows that it substantially not possible to simultaneously pre-code the channel input against two independent channel states.

\medskip
\noindent
$\bullet$ {\bf  Writing onto Fast Fading Dirt (WFFD) channel.}
For the WFFD channel with Gaussian fading, the authors of \cite{bennatan2008fading} optimize the achievable strategy in \eqref{eq:Capacity of GP channel}  over
 all jointly Gaussian distributions of $S,U$ and $X$.
\begin{thm}{\bf Achievability with jointly Gaussian signaling \cite[Sec. IV]{bennatan2008fading} \cite[Th. 1]{avner2010dirty}.}
\label{th:linear assigment}
Consider the WFFD channel for  $A^N \sim \iid  \Ncal(0,1)$ and let  $\rho=(\rho_{XS},\rho_{US},\rho_{UX})$ and
define $\mathbf{K}\subset [-1,1]^3$ as the region
\ea{
\mathbf{K}= \lcb \p{
|\rho_t|<1 \quad t \ \in\{ \rm XS,US,UX\} \\
1+2\rho_{XS} \rho_{US} - \rho_{XS}^2-\rho_{US}^2 - \rho_{UX}^2 =0,
}\rcb
\label{eq:linear assigment rho space}
}
 then an inner bound to capacity is
\ea{
\Ccal \geq R^{\rm IN} = \max_{\rho \in \mathbf{K}} \  \Ebb_{\theta} [ R_\Gamma(\rho, \theta) | \ A=\theta],
}
for
\ea{
 R_\Gamma(\rho,\theta) & = \f 12 \log \lb (P+c^2+ 2 \theta \rho_{XS} c \sqrt{P}+1)(1-\rho_{US}^2) \rb  \nonumber  \\
& \quad  -\f 12 \log \lb  P (1-\rho_{UX}^2)+c^2 (1-\rho_{US}^2) +2 \theta c(\rho_{XS}- (\rho_{UX} \rho_{US}))\sqrt{P}+1\rb.
}
\end{thm}
%
Th. \ref{th:linear assigment} attempts to generalize the result of \cite{costa1983writing} to the Gaussian fast fading case although, in all likelihood, one needs to consider a wider class of distributions than the jointly Gaussian distributions to attain maximum in \eqref{eq:Capacity of GP channel}.

\section{WFFD channel with a  discrete fading distribution}
\label{sec:Discrete fading distribution}
\subsubsection{Antipodal fading}
We begin by providing the approximate capacity for the WFFD channel in which the fading is uniformly distributed over the set $\{-1,+1\}$.
This is perhaps the simplest choice of fading distribution for this model, yet this example  well illustrates the main bounding techniques necessary to characterize the capacity.
\begin{thm}{\bf Outer bound and approximate capacity of  the WFFD channel with antipodal uniform fading.}
\label{th:Outer Bound for the Quasi-Static Fading Dirt Channel}
Consider the WFFD channel in which $A$ is uniformly distributed over the set $\{-1,+1\}$, then the capacity $\Ccal$ is upper bounded as
\ea{
\Ccal \leq R^{\rm OUT}  =
& \lcb \p{
 \f 12 \log(P+1)+\f 1 2
 &  c^2 \leq 1 \\
\f 12 \log(P+c^2+1) \\
\quad  - \f 14 \log(c^2)-\f1 2
& 1 < c^2 < P+1  \\
\f 1 4 \log(P+1)-\f 12
& c^2 \geq P+1,
} \rnone
\label{eq:outer bound antipodal}
}
and the capacity is to within $1 \ \bpcu$ from the outer bound in \eqref{eq:outer bound antipodal}.
\end{thm}
\begin{IEEEproof}
The achievability proof relies on a simple combination of superposition coding and DPC.
In the converse proof we define a ``conjugate'' sequences to the fading realization $A^N=a^n$, $\ao^N(a^N)=-a^N$,
and exploit  the fact that correct decoding must occur whether $A^N=a^N$ or $A^N=\ao^N$.

\noindent
$\bullet$ {\bf Achievability.}
Consider the achievable strategy in which the channel input is obtained as the superposition of two codewords:
(i) the codeword $X_{\rm SAN}^N$ (for \emph{State As Noise}), at rate $R_{\rm SAN}$, which treats $c A^N \circ S^N$ as additional noise and (ii)  the codeword $U_{\rm PAS}^N$ (for \emph{Pre-coded Against the State}), at rate $R_{\rm PAS}$,
which is pre-coded against $S^N$ as in the WDP channel.
This strategy attains the rate $R^{\rm IN}=R_{\rm SAN}+R_{\rm PAS}$ for
\ea{
R_{\rm SAN} & \leq   I(Y;X_{\rm SAN}|A) \nonumber \\
R_{\rm PAS} & \leq I(Y; U_{\rm PAS}|X_{\rm SAN},A) - I(U_{\rm PAS};S).
\label{eq:achievable rate noise and pre-coding}
}
Through the assignment
\ea{
& X_{\rm SAN}  \sim \Ncal(0, \al P),
\quad X_{\rm PAS}  \sim \Ncal(0, \alb P),  \quad X_{\rm SAN}\perp X_{\rm PAS}
\nonumber
 \\
& X  =X_{\rm SAN}+X_{\rm PAS},
\quad U_{\rm PAS} = X_{\rm PAS}+c \f{\alb P}{\alb P+1}  S,
\label{eq:assigment inner bound 2}}
for any $\al \in [0,1]$ and $\alb=1-\al$.
By further bounding  the expressions in \eqref{eq:achievable rate noise and pre-coding} for the assignment in \eqref{eq:assigment inner bound 2}, we obtain the achievable rate
\ea{
R^{\rm IN} (\al)\geq  \f 12 \log \lb 1 +  \f {\al P}{1 + \alb P + c^2} \rb 
+ \f 14 \log (\alb P+1) - \f 12.
\label{eq:Time Sharing + Binning Inner Bound app}
}
Optimizing the expression in \eqref{eq:Time Sharing + Binning Inner Bound app} over $\al$ results in  the inner bound
\ea{
 R^{\rm IN} =
&\lcb \p{
\f 12 \log \lb 1  + P \rb-\f 12                    & c^2 \leq 1  \\
\f 12 \log \lb 1+ P+ c^2 \rb  \\
\quad - \f 14 \log (c^2)-1  & 1< c^2 < P+1 \\
\f 1 4 \log \lb 1+ P\rb   -1                          & c^2\geq  P+1.
}
\rnone
\label{eq:inner bound two values}
}
\noindent
$\bullet$ {\bf Converse.}
Fano's inequality yields the upper bound
\ea{
 N(R-\ep_N) & \leq I(Y^N; W| A^N)
 \nonumber \\
& \leq \sum_{j=1}^N  H(Y_j|A_j)  - H(Y^N|A^N,W)
\nonumber \\
& \leq N \max_j H(Y_j|A_j)  - H(Y^N|A^N,W)
\nonumber \\
& \leq N \max_{P_{Y|A}} H(Y | A)- H(Y^N|A^N,W).
\label{eq:fano RCSI}
}
The  entropy term $\max_{P_{Y|A}} H(Y | A)$ in \eqref{eq:fano RCSI} is further bounded as
\ea{
 \max_{P_{Y|A}} H(Y| A)
& \leq \max_{P_{Y|A}}  \f 1 2 \lb H(X +c S + Z) + H (X-c S + Z_{j})\rb
\nonumber  \\
& \leq \max_{|\rho_{XS} | \leq  1}  \f 1 2 \lb H(X_{Gj} +c S + Z) + H (X_{Gj}-c S + Z_{j})\rb,
\label{eq:last eq positive enropy}
}
where \eqref{eq:last eq positive enropy} follows from the Gaussian Maximizes Entropy (GME)
property by letting $X_{Gj}$ be jointly Gaussian random variables with variance $P$ and with correlation $\rho_{XS}$ with $S$.
%
Optimizing \eqref{eq:last eq positive enropy} over $\rho_{XS}$ yields the upper bound
\ea{
 \max_{P_{Y|A}} H(Y | A)
& \leq  \f 12 \log (2 \pi e)^2 \lb P + c^2  + 1\rb,
\label{eq:positive ent 1}
}
where the maximum in \eqref{eq:last eq positive enropy} is attained for $\rho_{XS}=0$.
Define now ${\ao}^N(a^N)=-a^N$ and notice that
\ea{
H(Y^N|W,A^N)= \f  12 \sum_{a^N \in \{-1,+1\}^N}  \f 1 {2^N} \lb  H(Y^N|W,A^N=a^N)+ H(Y^N|W,A^N=-a^N)\rb,
}
so that
\eas{
- H(Y^N|W,A^N)
& \leq- \f 1 {2^{N+1}} \sum_{ a^N \in \{-1,+1\}^N}  \nonumber \\
& \quad \quad   H(X^N+c a^N \circ S^N+Z^N, X^N+c a^N \circ  S^N+Z^N  | W )   \nonumber  \\
& = - \f 1 {2^{N+1}} \sum_{ a^N \in \{-1,+1\}^N} \nonumber \\
 & \quad \quad    H(2 c a^N  \circ  S^N, X^N+c a^N \circ  S^N+Z^N|W)
\label{eq:transformation} \\
& = - \f 1 {2^{N+1}} \sum_{ a^N \in \{-1,+1\}^N} \lb  H(2 c a^N \circ  S^N|W) \rnone \nonumber \\
 & \quad \quad  \lnone +  H ( X^N+c a^N \circ  S^N+Z^N | S^N,W) \rb
\label{eq:transformation 2}\\
& = - \f 1 {2^{N+1}} \sum_{ a^N \in \{-1,+1\}^N}  \lb  H(2 c a^N\circ S^N)  \rnone \nonumber \\
 & \quad \quad  \lnone  +  H ( X^N+c a^N \circ  S^N+Z^N |S^N,W,X^N) \rb
\label{eq:transformation 3} \\
& = - \f 1 {2^{N+1}} \sum_{ a^N \in \{-1,+1\}^N}  H(2 c a^N\circ  S^N)+H(Z^N),
\label{eq:transformation 4}
}{\label{eq:antipodal}}
where \eqref{eq:transformation} follows from the fact that  transformation
\ea{
\lsb  \p {
T_{1i} \\
T_{2i} }
\rsb
= \lsb \p{
+1 & -1 \\
1 & 0
}\rsb  \cdot \lsb \p{
X_i + c a_i S_i +Z_i \\
X_i - c a_i S_i +Z_i
}\rsb,
\label{eq: transformation matrix}
}
 has unitary Jacobian. The equality in \eqref{eq:transformation 2} follows from the fact that $W \perp S^N$, \eqref{eq:transformation 3} from the fact that $X^N$ is a function of $W$ and $S^N$ and from the Markov chain $W-[X^N,S^N]-Y^N$.
Next, we observe that the terms in the summation in the RHS of \eqref{eq:transformation 4}  are all identical and equal to $1/2 \log ( 2 \pi e 4 c^2)+1/2 \log ( 2 \pi e)$,
so that
\ea{
-H(Y^N|W,A^N)
 \leq -  \f N4  \log ( 2 \pi e  c^2)- \f N 4 \log ( 2 \pi e)- \f N 2.
\label{eq:bound last}
}
Using \eqref{eq:positive ent 1} and \eqref{eq:bound last} we  rewrite  the outer bound in \eqref{eq:fano RCSI}  as
\ea{
R^{\rm OUT} & = \f 12 \log (2 \pi e )^2 \lb P + c^2  + 1\rb
\nonumber \\
& \quad \quad -  \f 14  \log ( 2 \pi e  c^2)- \f 1 4 \log ( 2 \pi e)- \f 1 2
\nonumber \\
 & = \f 12 \log (P+c^2+1)-\f 1 4 \log ( c^2)-\f 12.
 \label{eq:Outer Bound for the Quasi-Static Fading Dirt Channel}
}
Note that, as a function of $c^2$, the expression in  \eqref{eq:Outer Bound for the Quasi-Static Fading Dirt Channel} has a minimum in $c^2=P+1$.
From Lem. \ref{lem:capacity decreasing c} we have the capacity is decreasing in $c$: for this reason,
the channel in which $c^2$  is equal to $\min\{c^2,P+1\}$ corresponds to a model with larger capacity.
For this latter model, the outer bound in \eqref{eq:Outer Bound for the Quasi-Static Fading Dirt Channel} still holds so that
letting  $c^2$ equal to $\min\{c^2,P+1\}$ in \eqref{eq:Outer Bound for the Quasi-Static Fading Dirt Channel}  provides an outer bound to the capacity of the original model.
With this substitution and some further bounding for the case $c^2<1$, we obtain the outer bound in \eqref{eq:outer bound antipodal}.
By comparing the outer bound  in \eqref{eq:outer bound antipodal} and the inner bound  in \eqref{eq:inner bound two values}, we verify that the they differ of at most $1 \ \bpcu$.
\end{IEEEproof}
The result in Th. \ref{th:Outer Bound for the Quasi-Static Fading Dirt Channel} is conceptually simple but provides insights on more general scenarios.
The parameter $c$ controls the variance of the fading-time-state term:
(i) for small values of $c$, treating the term $cA^N\circ  S^N$ as additional noise results in a limited rate loss.
(ii) when the variance of $cA^N\circ S^N$ is larger than the transmit power, then it is approximately optimal to pre-code against one
fading realization,
as this strategy
grants correct decoding for half of the channel uses on average. Finally, (iii) in the intermediate regime capacity is approached by a combination of the previous two strategies.

\begin{rem}
The approximate capacity result for the two-user CCDP channel with independent, equal-variance states in
Th. \ref{th:Outer bound for the two-user, independent state case } has interesting similarities to the proof of Th. \ref{th:Outer Bound for the Quasi-Static Fading Dirt Channel} and the approximate capacity expressions in \eqref{eq:outer lapidoth} and \eqref{eq:outer bound antipodal} are also similar.
The achievability proof for both Th. \ref{th:linear assigment}  and Th. \ref{th:Outer Bound for the Quasi-Static Fading Dirt Channel} relies on a combination of superposition coding and DPC while, for the converse proof, the outer bound is tightened by using the fact that capacity is decreasing in the parameter $c$.
Despite these similarities, the two channel models are fundamentally different: from a high level perspective, in the WFFD channel each fading realization can be thought of as a compound user in the CCDP channel so that the number of compound user grows with the transmission length, instead of being constant.
\end{rem}
\subsubsection{WFFD  channel with a discrete fading distribution with mode larger than half}
\begin{thm}{\bf Outer bound and approximate capacity for the WFFD channel with a fading distribution of mode larger than half.}
\label{thm:approximate capacity discrete mass>1/2}
Consider the WFFD  channel in which $P_A(a)$ is a discrete distribution such that
\ea{
\exists \ m \in \Acal, \ \ST  P_A(m) \geq  \f 12,
\label{eq:cond mode}
}
and let $Q_m=P_A(m)$ and $\Qo_m=1-Q_m$,
then the capacity $\Ccal$ is upper bounded as
\ea{
& \Ccal \leq R^{\rm OUT}=  \label{eq:outer bound >1/2} \\
&  \lcb
\p{
\f 12 \log(1+P)+1 & \Qo_m \geq Q_m c^2 (1+\mu_A^2)\\
\f 1 2 \log(1+P) &  \Qo_m < Q_m c^2 (1+\mu_A^2)  \leq \Qo_m(P+1)  \\
 \ - \f {\Qo_m}2 \log \lb  {c^2 (1+\mu_A^2)} \rb + G_m  \\
 \f {Q_m} 2 \log(1+P)   + G_ m & Q_m  c^2(1+\mu_A^2) > \Qo_m(P+1),
} \rnone
\nonumber
}
for
\ea{
 G_m =\f 1 2    \Ebb_{A}\lsb \log \lb \f{1+\mu_A^2} { (A-m)^2} \rb |A \neq m \rsb +3,
\label{eq:inner outer gap mode}
}
and the capacity is to within a gap of
\ea{
G_m'= \f 1 2\Ebb_A \lsb \log \lb  (1 + \mu_A^2)  \lb \f 1 {A^2}+\f 1 {(A-m)^2} \rb \rb | A \neq m  \rsb+3,
\label{eq:inner outer gap mode 2}
}
from the outer bound in \eqref{eq:outer bound >1/2}.
%
\end{thm}
\begin{IEEEproof}
For the class of fading distributions in \eqref{eq:cond mode},  state pre-cancellation can be attained for a portion $Q_m$ of the channel uses on average:
for this reason, the achievable strategy employed in the proof of  Th. \ref{thm:approximate capacity discrete mass>1/2} is still effective.
In the converse proof, we extend the idea of conjugate fading sequences in the proof of  Th. \ref{th:Outer Bound for the Quasi-Static Fading Dirt Channel} to the elements in the set of  typical fading realizations.
The full proof can be found in App. \ref{app:approximate capacity discrete mass>1/2}.
\end{IEEEproof}
%
The next lemma provides a simplification of the result in  Th. \ref{th:Outer bound for the two-user, independent state case } under some conditions
on the support of $A$.
\begin{lem}
\label{lem:bigger gap}
If $|A|>\De$ and $|A-m| \geq \De$ for some $\De>0$, then \eqref{eq:inner outer gap mode} and \eqref{eq:inner outer gap mode 2} satisfy
\ean{
 G_m &  \leq \f{\Qo_m}{2} \log \lb \f {1+\mu_A^2} {\De^2}\rb +3  \\
 G_m' & \leq G_m+\f 1 2.
}
%
\end{lem}
\begin{IEEEproof}
The proof is omitted for brevity.
\end{IEEEproof}
Lem. \ref{lem:bigger gap} shows that a tight characterization of capacity is possible when the mean of $A$ is small
and the points in the support  are sufficiently  far from the mode of the distribution.

As an example of the result in Lem. \ref{lem:bigger gap}, consider the case in which $A$ has geometric distribution, i.e.
\ean{
P[A=k\De]=\qo^k q,
}
for $k\in \Nbb$, $q \geq 1/2$ and $\De=q/\sqrt{\qo}$ to obtain unitary variance:
in this case, we have
$G_m  \leq 3.15$ and $G'_m \leq 3.65$.
%

\subsubsection{WFFD channel in the ``strong fading'' regime}

\begin{thm}{\bf Outer bound and approximate capacity for the WFFD  channel in the  ``strong fading'' regime.}
\label{thm:approximate capacity strong}
Consider the WFFD channel with $c>2$ and in which $A$  is uniformly distributed over a discrete set
$\Acal=\lcb \al_i \rcb_{i=1}^M, \ \al_1 < \al_2 < \ldots < \al_M$
such that
\eas{
\al_1 & \geq \f 1 {c-1},
\label{eq:strong fading intro 1}\\
 \quad \al_{i+1} &  \geq c \al_{i}, \quad  i \in [2 \ldots M-1],
}{\label{eq:strong fading intro}}
then, the capacity $\Ccal$ is upper bounded as
\ea{
\Ccal \leq R^{\rm OUT} = \lcb \p{
\f 12 \log \lb 1+P \rb+G_s  & \f {1} M \geq \f {(1+\mu_A^2) c^2} M\\
 \f 1 {2} \log(1+P+(1+\mu_A^2) c^2) & \f {1} M < \f {(1+\mu_A^2) c^2} M \leq \f {M-1}{M}(P+1) \\
 \quad -\f {M-1}{2 M } \log \lb {(1+\mu_A^2) c^2} \rb+G_s &  \\
 \f 1 {2M} \log(1+P)  +G_s & \f {(1+\mu_A^2) c^2} M > \f {M-1}{M}(P+1),
} \rnone
\label{eq:outer strong}
}
for
\ea{
G_s & = \f 1 2 \log(1+\mu_A^2)+\f 1 2,
\label{eq:outer bound gap strong}
}
and the capacity is to within  a gap of $ G_s+1$
from the outer bound in \eqref{eq:outer strong}.
\end{thm}

\begin{IEEEproof}
As for Th. \ref{th:Outer Bound for the Quasi-Static Fading Dirt Channel}, the achievability proof  relies on the simple combination of superposition coding and DPC.
The converse bound involves defining $M-1$ conjugate sequences which are used to recursively bound the channel capacity by also providing a carefully-chosen genie-aided side information.
The proof is provided in App. \ref{app:approximate capacity strong}.
\end{IEEEproof}
%
%
%

\begin{rem}
The result in Th. \ref{thm:approximate capacity strong} can be generalized to the case in which $|a_{i+1} |\geq \kappa c| a_{i}|$ for some $\ka\in \Rbb^+$: in this case, \eqref{eq:outer bound gap strong} is expressed
\ea{
G_s & = \f 1 2 \log \ka (1+\mu_A^2)+\f 1 2.
}
%
\end{rem}
As an example of the result in Th. \ref{thm:approximate capacity strong}, consider the case in which $A$ is uniformly distributed over the set
\ean{
\Acal(M)= \lcb  \Delta,c  \Delta, c^2 \Delta, \ldots, c^{M-1} \Delta \rcb,
}
for $M\geq 3$ where $\Delta$ is chosen so as to obtain unitary variance \footnote{
More specifically, let
$\Delta = \f 1 {\sqrt{V}}$ with
$ V      = \f 1 M \f{1-c^{2 M}}{1-c^2}-\lb \f 1 M \f{(1-c^{M})}{(1-c)} \rb ^2$.
}, then $G_s  = \f 3 2$, $G_s' =  \f 5 2$.
%
Note that the capacity goes to zero when both $c$ and $M$ grow to infinity.
%
\section{WFFD channel with a continuous fading distribution}
\label{sec:WFFD with continuos fading}
The results derived in the previous section are limited to the case of discrete fading distributions: although relevant from a theoretical standpoint, this scenario is not particularly  meaningful in practical applications.
In this section, we show how the case of a discrete fading distribution aids the study of a continuous fading distribution.
In particular, we show that an outer bound to the capacity of the WFFD channel with a continuous fading distribution can be derived  by
considering the model with a discrete fading distribution obtained by quantizing the continuous one.
%
%

A first upper bound on the capacity of the WFFD channel with continuous fading can be obtained by adapting the result in
Th. \ref{th:Outer Bound for the Quasi-Static Fading Dirt Channel} to the case of a symmetric distribution.
\begin{lem}{\bf Outer bound for symmetric fading distribution.}
\label{lem:Outer bound for symmetric fading distribution}
Consider the WFFD channel in which $P_A(a)$ is continuous and symmetric, then the expression in \eqref{eq:outer bound antipodal} is an outer bound to capacity.
\end{lem}
\begin{IEEEproof}
The converse proof of Th. \ref{th:Outer Bound for the Quasi-Static Fading Dirt Channel} can be adapted to the case of continuous
distributions by similarly defining a conjugate sequence $\ao^N(a^N)=-a^N$.
Since the fading distribution is symmetric, the two sequences have the same probability and
 the bounding in \eqref{eq:antipodal} can be repeated for this  class of distributions.
\end{IEEEproof}
Generally speaking,  Lem. \ref{lem:Outer bound for symmetric fading distribution} only provides a loose upper bound to capacity
as it does not depend on the fading distribution; nonetheless, Lem. \ref{lem:Outer bound for symmetric fading distribution} shows
relatively simple conditions under which the capacity of the WFFD  channel is at most half of the AWGN capacity.
Unfortunately, the converse proof  of Th. \ref{th:Outer Bound for the Quasi-Static Fading Dirt Channel} does not extend to the
 continuous case
  unless the fading distribution
 concentrates  around the origin.
The next theorem extends the result in Th. \ref{thm:approximate capacity discrete mass>1/2} to the case of continuous fading distributions.
\begin{thm}{\bf Outer bound and approximate capacity for narrow fading.}
\label{th:narrow fading}
Consider the WFFD channel with $c>1$ in which $P_A(a)$  is a continuous distribution with
\ea{
\Pr \lsb \labs A-\mu_A \rabs \leq  \f 1 c \rsb =Q_m \geq \f 1 2,
\label{eq:narrow conditions}
}
then the expression in \eqref{eq:outer bound >1/2} for
\ea{
 G_m &  \leq \f{\Qo_m}{2} \log \lb 1+\mu_A^2 \rb +4,
\label{eq:bigger gap}
}
is an outer bound to capacity and the capacity is to within a gap of $G_m+ 1/2 \ \bpcu$ from the outer bound in \eqref{eq:outer bound >1/2}.
\end{thm}
\begin{IEEEproof}
The achievability proof follows a similar derivation as the
achievability proof of Th. \ref{thm:approximate capacity discrete mass>1/2}.
The converse proof is obtained in two steps: (i) first it is shown  that the capacity of the channel with
continuous fading distribution $A$ is to within a constant gap from the capacity of the channel with discrete fading distribution $A_{\De}$
when $A_{\De}$ is obtained by uniformly quantizing $A$, then (ii)
the result in Lem. \ref{lem:bigger gap} is applied to the model with  fading distribution $A^{\rm \De}$ to
obtain the approximate capacity of this model.
In the following, we prove step (i) while only an outline of the proof of step (ii) is provided for brevity.

\noindent
$\bullet$ {\bf Gap from capacity.}
Let the random variable $A_{\Delta}$ be defined as
\ea{
& \Pr[A_{\De}(A) =\gamma_k]=\Pr \lsb  A \in I_k    \rsb  \nonumber \\
&  I_k  =\lsb \mu_A+k \De- \f \De 2, \mu_A+(k+1)\De+\f \De 2  \rsb \nonumber\\
&  \gamma_k  =\Ebb[ A| A \in I_k],
\label{eq:def De}
}
for $k \in \Zbb$ and some $\De \in \Rbb^+$, that is, $A_{\Delta}$ is obtained by uniformly quantizing $A$ with step size $\De$ and so that
$\Ebb[A]=\Ebb[A_{\De}]$.
Next, define
\ea{
E^N & =c(A^N-A_{\De}^N)\circ S^N+Z^N-Z_{\De}^N,
\label{eq:side info}
}
for $Z_{\De}^N \ \sim i.i.d. \ \Ncal(0,1)$.
An outer bound to capacity can be obtained by providing $E^N$ in \eqref{eq:side info} to the receiver as a genie-aided side-information, that is
\eas{
N(R-\ep_N)
& \leq I(Y^N,E^N ;W|A^N) \nonumber \\
& = I(Y^N-E^N,E^N ;W|A^N)
\label{eq:def Yo1 }\\
& = I(Y_{\De}^N;W|A^N) + I(E^N ;W|A^N,Y_{\De}^N),
\label{eq:def Yo 2}
}{\label{eq:def Yo}}
where \eqref{eq:def Yo1 } follows from the fact that the transformation has unitary Jacobian while in  \eqref{eq:def Yo 2} follows
by  defining
$Y_{\De}^N=X^N + c A_{\De}^N \circ S^N+Z_{\De}^N$.
The term $I(E^N ;W|A^N,Y_{\Delta}^N)$ is further bounded as
\ean{
I(E^N ;W|A^N,Y_{\Delta}^N) &= H(E^N|A^N,Y_{\Delta}^N) - H(E^N|A^N,Y_{\Delta}^N,W)  \\
& = H(E^N|A^N,Y_{\Delta}^N) - H(Z^N|A^N,\Zo^N,W,S^N,X^N) \\
& = H(E^N|A^N) - \f N 2 \log 2 \pi e \\
& \leq N \max_i H(E_i|A_i) - \f N 2 \log 2 \pi e \\
& \leq N \max_{P_{E|A}} H(E|A) - \f N 2 \log 2 \pi e.
}
Note that $A_{\Delta}$ is a deterministic function of $A$, so that the entropy term $H(E|A)$ can be bounded as
\ean{
 H(E|A)
 & \leq  \int_{\Acal}  \f 12 \log 2 \pi e  \lb c^2 (a-A_{\Delta}(a))^2+2 \rb \diff P_A
   \leq \f 12 \log \lb c^2 \De^2+2 \rb.
\label{eq: last integral}
}
From \eqref{eq: last integral} we conclude that, by choosing  $\De=1/c$  in \eqref{eq:def De},  the capacity of the WFFD channel with fading distribution $P_{A_{\De}}$  is to within a gap of $1 \ \bpcu$ from the capacity of the channel with fading  distribution $P_A(a)$.
When the condition in \eqref{eq:narrow conditions} holds, the mode of $A_{\De}$ is $A_{\De}=\gamma_0 \in [\mu_A-1/c,\mu_A+1/c]$  with $P_{A_{\De}}(\gamma_0) \geq 1/2$ and thus the result in Th. \ref{thm:approximate capacity discrete mass>1/2} can
be applied.
Note also that the distribution of $A_{\De}$ does not necessarily have unitary variance, so that Lem. \ref{lem:Mean and variance of the fading-time-state term}  can be invoked to normalize the fading variance.
This normalization, though, does not affect the outer bound expression.
\end{IEEEproof}
The result in Th. \ref{th:narrow fading} is analogous to the result in Th.  \ref{thm:approximate capacity discrete mass>1/2}
as it identifies the condition under which it is approximately optimal
for the transmitter  to pre-code against one realization of the fading distribution
 while treating the remaining randomness in the fading as noise.
%
%

\begin{rem}
Note that the condition in \eqref{eq:narrow conditions} can be generalized to
\ea{
\Pr \lsb \labs A-m \rabs \leq  \f \ka c \rsb =Q_m > \f 1 2,
}
for some value $m \in \Acal$ to obtain a more general result than Th. \ref{th:narrow fading}.
This yield an expression for $G_m$ as in \eqref{eq:inner outer gap mode 2} and a gap from capacity as in \eqref{eq:inner outer gap mode 2}.
\end{rem}

The next theorem shows that  there exists a class of fading
distributions for which, regardless of the available transmit power,  the capacity of the WFFD channel substantially reduces the capacity of the channel without transmitter state knowledge.
We denote  the indicator function for the set $x \in I$ as $1_{\{x \in \Xcal\}}$.

\begin{thm}{\bf An example with a fat-tailed distribution.}
\label{th:fat-tailed distribution}
Consider a WFFD channel with $c>2$, then there exists a distribution of the form
\ea{
P_A(a)=\f {\al} a \cdot 1_{\{a \in I\}},
\label{eq:fat tail}
}
such  that
capacity is upper bounded as
\ea{
R^{\rm OUT} & = \f 12 \log \lb 1  + \f {P}{1+c^2}\rb +2,
\label{eq:outer fat}
}
and for which capacity is to within $3 \ \bpcu$ from the outer bound in \eqref{eq:outer fat}.
\end{thm}
\begin{IEEEproof}
Quantizing the distribution in \eqref{eq:fat tail} as in Th. \ref{th:narrow fading} in intervals of size $[c^{-k},c^{-(k-1)}]$
yields a random variable $A_{\De}$  which satisfies the conditions of  Th. \ref{thm:approximate capacity strong}.
The support $I$ can be chosen as $[\ka c^{-M-1}, \ka c^{-1}]$ for some sufficiently large $M$ so that  $(1+\mu_A^2) c^2/ M \leq (P+1)(M-1)/M$,
thus yielding the outer bound in \eqref{eq:outer fat} while $\mu_A \leq 1$.
The achievability proof follows by treating the fading-times-state term as noise.
The full proof is omitted for brevity.
\end{IEEEproof}
%
%
\subsubsection*{Discussion}
Unfortunately, we are currently unable to determine a characterization of capacity for continuous fading distributions of practical
relevance, such as Gaussian, Rayleigh or uniform  distributions.
Also, we are unable to determine the asymptotic behaviour  of capacity as $c$  grows large:
in this regime, one would expect state pre-coding to become ineffective as in Th. \ref{th:fat-tailed distribution}.
For the case of zero mean fading this implies, in particular, that state knowledge at the transmitter does not provide
any substantial rate advantage with respect to the channel without transmitter
state knowledge.
In practical systems, state knowledge at the transmitter often come at the cost of an increase in complexity in the network architecture and
transmitter design:
as such, determining the fading regimes in which transmitter knowledge is rendered useless by the presence of
fading is of great practical interest.
\section{Conclusions}
\label{sec:Conclusion}
This paper investigates the capacity of the Writing of Fast Fading Dirt (WFFD) channel, a variation of the classic ``writing on dirty paper'' channel in which the state sequence is multiplied by an ergodic fading sequence known only at the receiver.
In this channel, then, the output is obtained as the sum of the  channel input, additive Gaussian noise and a fading-times-state term which is the element-wise product of the channel state, known only at the transmitter, and the fading process, known only at the receiver.
We focus on the case in which the channel state is a white Gaussian process and the fading sequence is at i.i.d. sequence with either a discrete or a continuous distribution.
The WFFD channel is a special case of the Gelf'and-Pinsker channel for which capacity is known:
unfortunately, capacity is expressed as a solution of a maximization problem that cannot be easily determined in closed-form or evaluated numerically.
For this reason, we derive alternative inner and outer bounds to capacity and bound their respective distance for certain classes of fading distributions.
For the WFFD channel with a discrete fading distribution, we determined capacity to within a small gap for two classes of distributions: distributions with  mode larger than a half and uniform distributions in which the points in the support are incrementally spaced apart.
For the WFFD channel with a continuous fading distribution, we derive capacity for the case in which more than half of the probability is concentrated in a small interval.
In all these cases, capacity is approached by letting the channel input be the superposition of two codewords: a codeword treating the fading-times-state as additional noise and a codeword pre-coded against one realization of the fading times the state sequence.
This relatively simple attainable strategy shows, from a high-level perspective, that robust state pre-cancellation is substantially unsuccessful for these fading distributions.

\appendices
%
\section{Proof of Lem. \ref{lem:capacity decreasing c}}
\label{app:capacity decreasing c}

Consider two sequences $S_{1}^N$ and $S_{2}^N$ such that $S_m^N \sim \iid \Ncal(0,Q_m), \ m \in \{1,2\}$, $S_1^N \perp S_2^N$
with $Q_1+Q_2=1$ and let the channel state of the WFFD  channel  be obtained as $S^N=S_{1}^N+S_{2}^N$.
%
Providing the sequence $S_{2}^N$ to both the transmitter and receiver can only increase capacity, since they can disregard this extra information and operate as in the original channel.
The channel in which $S_{2}^N$ is provided to both encoder  and decoder falls in the class of channels studied in \cite[Th. 1]{cover2002duality} for which capacity can be bounded as
\eas{
\Ccal
&=  \max_{X,U| S_{2},S_{R}} I(X+cS A + Z,  A, S_{2} ; U) -I(U; S, S_{2})
\nonumber \\
&=\max_{X,U| S_{2},S_{1}} I(X+cS_{1} A + Z,  U| A,S_{2}) -I(U; S_{1} | S_{2})
\nonumber \\
& \leq \max_{X,U| S_{2},S_{1}} I(X+cS_{1} A + Z; U,S_{2}|A) -I(U,S_{2}; S_{1})
\label{eq: independent S1 S2}\\
& = \max_{X,\Ut| S_{1}} I(X+cS_{1} A + Z; \Ut|A) -I(\Ut; S_{1}),
\label{eq: independent S1 S2 last}
}{\label{eq:genie}}
where,  \eqref{eq: independent S1 S2} follows from the independence of $S_{1}$ and $S_{2}$ by defining $\Ut=[U \ S_{2}]$ in \eqref{eq: independent S1 S2 last}.
Since $S_{2}$ no longer appears in \eqref{eq: independent S1 S2 last}, it can be dropped from the maximization.
From the result in Lem. \ref{lem:Mean and variance of the fading-time-state term}, we have that
\eqref{eq: independent S1 S2 last} equals the capacity of the channel in \eqref{eq:Dirty Paper Channel} for which $\ct  =c/\sqrt{Q_1}$ instead of $c$.
From this equivalence, we conclude that the capacity of the WFFD channel is decreasing in the parameter $c$.

\section{Proof of Th. \ref{thm:approximate capacity discrete mass>1/2}}
\label{app:approximate capacity discrete mass>1/2}
\noindent
$\bullet$ {\bf Achievability.}
%
Consider the achievable strategy in  Th. \ref{th:Outer Bound for the Quasi-Static Fading Dirt Channel} and let
the top codeword $U_{\rm PAS}$ in \eqref{eq:assigment inner bound 2}
be pre-coded against the sequence $c m S^N$ as in the WDP channel. 
%
This assignment attains the rate
\ean{
 R_{\rm PAS}
& =  \lsb I(Y;U|X^{\rm SAN},A)-I(U;S)\rsb^+ \\
%
& \geq \f { Q_m} 2  \log(1+\alb P)  \nonumber  \\
&  +\sum_{a \in \Acal \setminus \{m\}}  \f {P_A(a)}2 \log \lb \f {(1+c^2 a^2+\alb P)(1+\alb P)}{\alb P c^2 (a-m)^2+\alb P+c^2a^2+1} \rb \\
& \geq \f {Q_m} 2  \log(1+\alb P)  \nonumber  \\
&  -\sum_{a \in \Acal \setminus \{m\}}   \f { P_A(a)} 2 \log \lb \f{(a-m)^2}{a^2}+1  \rb,
}
while the overall attainable rate $R^{\rm IN}(\al)$ in \eqref{eq:Time Sharing + Binning Inner Bound app} becomes
\ea{
 R^{\rm IN}(\al) &  = \f 12  \Ebb_A  \lsb \log \lb 1 + \f {\al P}{1+c^2A^2+\alb P}\rb \rsb  \nonumber \\
 & + \f { Q_m} 2  \log(1+\alb P)  \nonumber  \\
&  -\sum_{a \in \Acal \setminus \{m\}}   \f { P_A(a)} 2 \log \lb \f{(a-m)^2}{a^2}+1  \rb.
\label{eq:inner tot}
}
The choice of  $\alb P$  in \eqref{eq:inner tot} as
\ea{
\alb^* P=\max  \lcb \min \lcb \f{Q_m}{\Qo_m} c^2(1+\mu_A^2)-1, P\rcb, 0 \rcb,
}
yields the inner bound
\ea{
& R^{\rm IN}=
\label{eq:inner bound >1/2} \\
& \lcb \p{
 \f 12 \log \lb 1 + \f{P} {1+c^2(1+\mu_A^2)} \rb & \Qo_m \geq Q_m c^2(1+\mu_A^2) \\
\f 12 \log (P+c^2 (1+\mu_A^2) +1)   &  \Qo_m \leq Q_m c^2 (1+\mu_A^2)  \leq \Qo_m(P+1) \\
\quad - \f {\Qo_m} 2 \log \lb c^2 (1+\mu_A^2)  \rb -G_m & \\
%
%
 \f {Q_m} 2 \log(1+P) -G_m  & Q_m c^2 (1+\mu_A^2) > \Qo_m(P+1),
}\rnone \nonumber
}
for
\ea{
G_m= \f {1}{2} \Ebb_A \lsb \log \lb \f {(A-m)^2} {A^2} +1 \rb |A \neq m \rsb+1,
}

\noindent
$\bullet$ {\bf Converse.}
Using Fano's, inequality we write
\eas{
  N(R- \ep_N)  &    \leq I(Y^N;W|A^N)  \nonumber \\
 & \leq N \max_j H(Y_j|A_j)  \nonumber \\
 & \quad  \quad \quad  -H(Y^N|W,A^N) \nonumber \\
 & \leq \f N2 \Ebb_A \lsb \log2 \pi e(P+A^2 c^2 +2 |c||A|\sqrt{P}  +1 )\rsb \nonumber \\
 & \quad  \quad \quad   -H(Y^N|W,A^N) \label{eq: GME here}  \\
%
%
& \leq \f N2 \log2 \pi e(P+c^2(1+\mu_A^2)  +1 ) \nonumber \\
 & \quad  \quad \quad -H(Y^N|W,A^N) +\f N2,
%
\label{eq:fano 1}
}{\label{eq:fano}}
where \eqref{eq: GME here} follows from the GME and \eqref{eq:fano 1} follows from Jensen's inequality.
Next, we derive a bound on the entropy term $H(Y^N|W,A^N)$ based on the
properties of  the set of typical fading realizations, $\Tcal_{\ep}^N(P_A)$, defined as
\ea{
& \Tcal_{\ep}^N(P_A)  = \label{eq:def strong typical} \\
& \lcb   a^N, \ \labs \f 1 N N(k|a^N)-P_A(k) \rabs \leq \ep P_A(k),  \quad \forall \ k \ \in \Acal \rcb,
\nonumber
}
where $N(k|a^N)$ is the number of symbols $k \in \Acal$ in the sequence $a^N$, that is
\ea{
N(k|a^N)= \sum_{i=1}^N 1_{\{a_i=k\}}.
}
%
For the typical set in \eqref{eq:def strong typical}, we have
\eas{
 P(a^N)  & \leq \f 1 {2^{n(1+\ep) H(A)}}, \quad a^N \in \Tcal_\ep^N
\label{eq:typical properties probability} \\
 \labs \Tcal_\ep^N(P_A) \rabs & \leq (1-\delta_{\ep})2^{N(1-\ep) H(A)}
\label{eq:typical properties cardinality}\\
 N(k|a^N) & \leq   N P_A(k)(a)(1-\ep),
}{\label{eq:typical properties} }
for $\delta_{\ep} = 2|\Acal|e^{ - N 2 \min_k P_A(k)}$.
When the block-length $N$ is sufficiently large, we have that  $\ep \leq  (Q_m-\f  12)/ Q_m$ in \eqref{eq:def strong typical} which implies $N(m|a^N)>1/2$.
For $N(m|a^N)>1/2$, there exists a one-to-one mapping $\ao^N(a^N): \ \Tcal_\ep^N(P_A) \goes \Tcal_\ep^N(P_A)$  such that
\ea{
{\rm if} \  \ao_i\neq m  \quad {\rm then} \ a_i=m \nonumber \\
{\rm if} \  a_i\neq m  \quad {\rm then} \ \ao_i=m,
\label{eq:properties mapping}
}
that is, the sequence $\ao^N(a^N)$ is obtained by permuting the $N-N(m|a^N)$  indexes for which $a_i \neq m$ with some
$N-N(m|a^N)$ indexes for which $a_i = m$, while $2 N(m|a^N) - N$ indexes are such that $a_i=\ao_i=m$.
Since the mapping $\ao^N(a^N)$ in \eqref{eq:properties mapping} is a one-to-one mapping of the typical set onto itself,  we must have
\ea{
\sum_{a^N \in \Tcal_\ep^N(P_A)} P(a^N) H(Y^N|W,A^N=a^N) = \sum_{ a^N \in \Tcal_\ep^N(P_A)} P\lb \ao^N(a^N) \rb  H\lb \Yo^N|W,A^N=\ao^N(a^N)\rb,
\label{eq:conj 2}
}
where $\Yo^N$ in \eqref{eq:conj 2} is defined as
\ea{
\Yo^N=X^N+c \ao^N S^N + \Zo^N,
\label{eq:Ao definition}
}
for $\Zo^N \sim i.i.d. \ \Ncal(0,1), \ \Zo^N \perp Z^N$.
Using the definitions above, we have that the entropy term $H(Y^N|W,A^N)$ can be bounded as
\eas{
- H(Y^N|W,A^N)
& = -\sum_{a^N \in \Acal^N } P(a^N) H(Y^N|W,A^N=a^N)  \nonumber \\
& \leq -\sum_{a^N \in \Tcal_\ep^N(P_A)} P(a^N) H(Y^N|W,A^N=a^N)
\nonumber \\
& = - \f12 \sum_{a^N \in \Tcal_\ep^N(P_A)} P(a^N) \lb H(Y^N|W,A^N=a^N) \rnone \\
& \quad \quad \lnone +H(Y^N|W,A^N=\ao^N) \rb
\nonumber  \\
& \leq - \f12 \sum_{a^N \in \Tcal_\ep^N(P_A)} P(a^N)
\nonumber  \\
& \quad \quad  \lb H(X^N+c a^N \circ S^N + Z^N,X^N+c \ao^N \circ S^N + \Zo^N |W) \rb
\nonumber  \\
& = - \f12 \sum_{a^N \in \Tcal_\ep^N(P_A)} P(a^N)
\nonumber  \\
& \quad H \lb c (a^N-\ao^N)\circ S^N + Z^N- \Zo^N,X^N+c \ao^N \circ S^N + \Zo^N |W \rb
\nonumber  \\
& = - \f12 \sum_{a^N \in \Tcal_\ep^N(P_A)} P(a^N) \lb  H \lb c (a^N-\ao^N)\circ S^N + Z^N- \Zo^N \rb+ \rnone
\nonumber \\
& \quad \quad \quad \quad \lnone  H(\Yo^N|Y^N-\Yo^N,W,S^N,X^N   ) \rb
\label{eq: trick larger 1/2 1} \\
& \leq - \f12 \sum_{a^N \in \Tcal_\ep^N(P_A)} P(a^N) \cdot  \nonumber \\
 & \lb  H \lb c (a^N-\ao^N)\circ S^N + Z^N- \Zo^N \rb + H(\Zo^N|\Zo^N-Z^N) \rb
\nonumber \\
& =- \f12 \sum_{a^N \in \Tcal_\ep^N(P_A)} P(a^N) \cdot  \nonumber \\
&   H \lb c (a^N-\ao^N)\circ S^N + Z^N- \Zo^N \rb +  \f  N  2 \log ( \pi e ),
\label{eq: trick larger 1/2 2}
}{\label{eq: trick larger 1/2}}
where \eqref{eq: trick larger 1/2 1}  follows from the fact that $S^N$ and $Z^N$ are independent from $W$.
%
We continue the series of inequalities in \eqref{eq: trick larger 1/2} by noting that
\eas{
& \f1 2  \sum_{a^N \in \Tcal_\ep^N(P_A)} P(a^N) H \lb c (a^N-\ao^N)\circ S^N + Z^N- \Zo^N \rb
\nonumber \\
& \leq  - \f 12 \f 1 {2^{n(1+\ep) H(A)}}  \cdot
\label{eq:typical one} \\
& \quad \quad \sum_{a^N \in \Tcal_\ep^N(P_A)}  H \lb c (a^N-\ao^N)\circ S^N + Z^N- \Zo^N \rb
\nonumber  \\
& \leq - \f 12 \f 1 {2^{n(1+\ep) H(A)}} \cdot
\nonumber  \\
& \quad \quad  \sum_{a^N \in \Tcal_\ep^N(P_A)}  \sum_{i=1}^N \lb  H \lb c (a_i-\ao_i)S_i + Z_i- \Zo_i  \rb \rb,
\label{eq: trick larger 1/2V1 1}
}{\label{eq:typicality pass}}
where \eqref{eq:typical one} follows from the bound in \eqref{eq:typical properties probability} while
%
\eqref{eq: trick larger 1/2V1 1} follows from the fact that $S^N$ and $Z^N$ are i.i.d. sequences.
%
From
the definition of the mapping $\ao^N(a^N)$, the sequence $a_i-\ao_i$ can take three types of values:  $m-k$, $k-m$ and $0$
where $k$ is any element of $\Acal\setminus \{m\}$.
%
More specifically, $a_i-\ao_i=m-k$ occurs $N(k|a^N)$ times, $a_i-\ao_i=k-m$ occurs $N(k|a^N)$  times for all $k \in \Acal$ while
$a_i-\ao_i=0$ occurs $2(N-N(m|a^N))$ times.
Using these observations, we  write
\eas{
 \eqref{eq: trick larger 1/2V1 1}
 & = - \f 12 \f 1 {2^{n(1+\ep) H(A)}}  \sum_{a^N \in \Tcal_\ep^N(P_A)}  \nonumber\\
 & \quad \quad    \cdot \lb \sum_{k \in \Acal \setminus \{m\} } 2  N(k|a^N) H \lb c (m-k)S_i + Z_i- \Zo_i  \rb \rnone
\nonumber\\
& \quad \quad \quad   \lnone + \f 12  (2 N(m|a^N) - N) H \lb Z_i- \Zo_i  \rb \rb \\
& = - \f 12 \f 1 {2^{n(1+\ep) H(A)}} \sum_{a^N \in \Tcal_\ep^N(P_A)} \nonumber \\
&  \lb \sum_{k \in \Acal \setminus \{m\} } 2  N(k|a^N) H \lb c (m-k)S_i+Z_i- \Zo_i\rb \rnone  \nonumber \\
& \quad \quad \lnone- \f 12 (2 N(m|a^N) - N)  \log ( 4 \pi e) \rb \\
& \leq   - \f 12 \f 1 {2^{n(1+\ep) H(A)}}  \sum_{a^N \in \Tcal_\ep^N(P_A)} \nonumber \\
&  \sum_{k \in \Acal \setminus \{m\} } 2  N(k|a^N) \f 12 \log 2 \pi e (c^2 (m-k)^2 +2)  \nonumber \\
%
& =  - \f 1 {2^{n(1+\ep) H(A)}}  (1-\delta_{\ep})2^{n(1-\ep) H(A)} \cdot
\label{eq:user cardinality} \\
&  \sum_{k \in \Acal \setminus \{m\} }  N(k|a^N) \f 12 \log 2 \pi e (c^2 (m-k)^2+2 )   \nonumber \\
%
& =  - \f 1 {2^{n(1+\ep) H(A)}}  (1-\delta_{\ep})2^{n(1-\ep) H(A)}(1-\ep) N  \cdot \nonumber \\
& \sum_{k \in \Acal \setminus \{m\} } P_A(k)(1-\ep)\f N 2 \log 2 \pi e (c^2 (m-k)^2+2 ),
\label{eq:ep pa}
}{\label{eq:typicality 2}}
where \eqref{eq:user cardinality} follows from the bound on the cardinality of the typical set in \eqref{eq:typical properties cardinality}
and   \eqref{eq:ep pa} from the definition in  \eqref{eq:def strong typical}.
For $N$ is sufficiently large, we  have
\eas{
 - H(Y^N|W,A^N)  
%
& \leq - \sum_{k \in \Acal \setminus \{m\} } P_A(k)\f N2 \log 2 \pi e \lb c^2 (m-k)^2 +2\rb
\nonumber  \\
& \quad \quad - \f {N \Qo_m } 2  \log ( 2 \pi e)  - \ep_{\rm all}
\label{eq:epsilon all 1}\\
& \leq - \f {N \Qo_m} 2 \log 2 \pi e c^2 - \f N 2 \Ebb_A[\log(A-m)^2|A \neq m]
\nonumber  \\
& \quad \quad - \f {N \Qo_m} 2  \log (2 \pi e)- \ep_{\rm all},
}{\label{eq:epsilon all}}
for some $\ep_{\rm all}$ that goes to zero as $N \goes \infty$.
Using the bound for \eqref{eq:epsilon all} in \eqref{eq:fano 1} and for some $\ep_{\rm all}$ sufficiently small, we obtain
the outer bound
\ea{
& R^{\rm OUT} =  \f12 \log \lb 2 \pi e(P+c^2(1+\mu_A^2)  +1 )\rb
\label{eq:outer before optimization} \\
& \quad - \f{\Qo_m} 2 \log  (c^2(1+\mu_A^2)) - \f 1 4 \log( \pi e) -\f 1 2 \Ebb_A[\log \lb  \f { (A-m)^2}{1+\mu_A^2} \rb |A \neq m]  +\f 12.
\nonumber
}
Using Lem. \ref{lem:capacity decreasing c}, we can consider the assignment
\ea{
\min \lcb \f {\Qo_m}{Q_m }(1+P),c^2(1+\mu_A) \rcb,
\label{eq:optimal c mode}
}
for the term  ${c^2}(1+\mu_A^2)$ in \eqref{eq:outer before optimization} which yields the expression in \eqref{eq:outer bound >1/2}.

The  gap between inner and outer bound of $G_m'$ can be obtained by comparing the  expressions in \eqref{eq:outer bound >1/2}
and \eqref{eq:inner bound >1/2}.

\section{Proof of Th. \ref{thm:approximate capacity strong}}
\label{app:approximate capacity strong}

In the following, we provide the proof for the result in Th.  \ref{thm:approximate capacity strong} for the case of $M=3$:
%
an outline of the proof for the case of any $M>3$ is provided at the end of this section while the full derivation is omitted for brevity.

The achievability proof is a variation of the achievability proof of Th. \ref{thm:approximate capacity discrete mass>1/2}
when letting the codeword $U_{\rm PAS}$
be pre-coded against the sequence $c \mu_A S^N$ as in the WDP channel.

\noindent
$\bullet$ {\bf Converse.}
For $\Acal=\{\al_1,\al_2,\al_3\}$ and $\al_1 \leq \al_2 \leq \al_3$, we define two conjugate sequences of $a^N$,
$\ao_{(1)}^N(a^N)$ and $\ao_{(2)}^N(a^N)$, as follows:

\noindent
$\star$ the portion of $a^N$ equal to $\al_1$, is equal to $\al_2$ in $\ao^N_{(1)}$ and equal to $\al_3$ in $\ao^N_{(2)}$,

\noindent
$\star$
the portion of $a^N$ equal to $\al_2$, is equal to  $\al_3$ in $\ao^N_{(1)}$ and equal to $\al_1$ in $\ao^N_{(2)}$, and

\noindent
$\star$
the portion of $a^N$ equal to $\al_3$, is equal to  $\al_1$ in $\ao^N_{(1)}$ and equal to $\al_2$ in $\ao^N_{(2)}$.
%
%
%
%

From the definition of the mapping, $\ao^N_{(1)}(a^N)$ and $\ao^N_{(2)}(a^N)$, we have that
\eas{
a^N \in \Tcal_\ep^N(P_A) \iff \ao_{(k)}^N(a^N) \in \Tcal_\ep^N(P_A), \quad k\in[1,2],
}
moreover
\ea{
\sum_{a^N \in \Tcal_\ep^N(P_A)} P_{A^N}(a^N)   H(Y^N|W,A^N=a^N) = \sum_{a^N \in \Tcal_\ep^N(P_A)} P_{A^N}(\ao_{(k)}^N(a^N))   H(Y^N_{(k)}|W,A^N=\ao_{(k)}^N(a^N)),
\label{eq:equal summation}
}
where $Y_{(k)}^N$ is defined similarly to \eqref{eq:Ao definition} as
\ea{
Y^N_{(k)}= X^N+c \ao_{(k)}^N S^N + Z_{(k)}^N,
\label{eq:equivalent Yk}
}
for $Z_{(k)}^N \sim i.i.d. \ \Ncal(0,1), \ k\in [1,2]$.
Similarly to \eqref{eq:fano}, Fano's inequality yields the bound
\ea{
  N(R- \ep_N)   & \leq \f N2 \log2 \pi e(P+c^2(1+\mu_A^2)  +1 ) \nonumber \\
 & \quad  \quad \quad -H(Y^N|W,A^N) +\f N2.
\label{eq:bound outer 1 strong}
}
Using the equivalence in \eqref{eq:equal summation}, the term $H(Y^N|W,A^N)$
can be rewritten as
\ean{
& -H(Y^N|W,A^N)
  = - \sum_{a^N \in \Tcal_\ep^N(P_A)} \f 1 {3^N}   H(Y^N|W,A^N=a^N) \\
&  = - \f 1 {3^{N+1}} \sum_{a^N \in \Tcal_\ep^N(P_A)}  \lb H(Y^N|W, A^N=a^N) + H(Y^N|W, A^N=a_{(1)}^N) + H(Y^N|W,A^N=a_{(2)}^N) \rb
}
%
For $a^N \in \Tcal_\ep^N(P_A)$, we have
\eas{
& - H(Y^N|W, A^N=a^N) - H(Y^N|W, A^N=a_{(1)}^N) - H(Y^N|W,A^N=a_{(2)}^N)
\nonumber\\
& \leq - H(Y^N,Y_{(1)}^N,Y_{(2)}^N|W, A^N=a^N)  \label{eq:define y()} \\
& = - H \lb \{ Y_i,Y_{(1),i},Y_{(2),i},\ \forall i   \  a_i=\al_1 \},  \{ Y_i,Y_{(1),i},Y_{(2),i}, \ \forall i   \  a_i=\al_2 \}, \rnone  \label{eq:example 3 v0} \\
& \quad \quad \quad \quad \lnone \{ Y_i,Y_{(1),i},Y_{(2),i}, \ \forall i   \ a_i=\al_3 \} | W, A^N=a^N \rb
\nonumber  \\
& = - H \lb \{ Y_i,Y_i-Y_{(2),i},Y_{(2),i}, \ \forall i  \  a_i=\al_1 \}, \rnone
\label{eq:example 3 v1}\\
& \quad \quad \quad \ \  \{Y_{(2),i}-Y_{i},Y_{(1),i},Y_{(2),i}, \ \forall i \   a_i=\al_2 \}, \nonumber \\
& \quad \quad \quad \  \lnone  \{Y_{i}, Y_{(1),i}-Y_{(2),i},Y_{(2),i}, \ \forall i \     a_i=\al_3 \} | W, A^N=a^N \rb,
\nonumber
}{\label{eq:example 3}}
where
\eqref{eq:example 3 v0} follows by re-arranging the channel outputs
 according to the fading realization and \eqref{eq:example 3 v1} follows from the fact this the transformation has unitary Jacobian.
Consider the set
\ea{
& \{ Y_i-Y_{(1),i}, \   \forall i \  a_i=\al_1 \} \nonumber \\
& \quad \cup \{ Y_{(2),i}-Y_{i}, \ \forall i \   a_i=\al_2 \} \nonumber \\
& \quad \quad \cup \{ Y_{(1),i}-Y_{(2),i}, \ \forall i \   a_i=\al_3 \},
\label{eq:set minus one}
}
from the definition of the conjugate sequences, we have that the set in \eqref{eq:set minus one} contains the elements of the vector
\ea{
c (\al_2-\al_1)S^N+ \Zt_{21}^N,
\label{eq:set 1}
}
where  $\Zt_{21}^N$  is
obtained as
\ea{
{\Zt_{21,i}} = \lcb \p{
 &  Z_i-Z_{(1),i}   &  a_i=\al_1  \\
 &  Z_{(2),i}-Z_{i} &  a_i=\al_2  \\
 &  Z_{(1),i}-Z_{(2),i} &   a_i=\al_3.
 } \rnone
\label{eq:Z21}
}
Next, continuing the series of inequalities in \eqref{eq:example 3}, we have
\eas{
& - 3 H(Y^N|W,A^N=a^N)   \nonumber \\
& \leq -H \lb \{ Y_{i},Y_{(2),i}, \  a_i=\al_1 \}, \{ Y_{(1),i},Y_{(2),i}, \  a_i=\al_2 \}, \{ Y_{i},Y_{(1),i} \  a_i=\al_3 \} | W, c (\al_2-\al_1)S^N+\Zt_{21}^N, A^N=a^N \rb
\label{eq:onejmb} \\
& \quad  \quad - H(c (\al_2-\al_1)S^N+\Zt_{21}^N|W)\nonumber  \\
& \leq -H \lb \{ Y_i-Y_{(2),i},Y_i, \  a_i=\al_1 \}, \{ Y_{(2),i}-Y_{(1),i},Y_{(2),i} \  a_i=\al_2 \}, \{ Y_{(1),i}-Y_{i},Y_{(1),i} \  a_i=\al_3 \} \rnone \nonumber \\
& \quad \quad \lnone | W, c (\al_2-\al_1)S^N+\Zt_{21}^N, A^N=a^N \rb  \nonumber \\
& \quad  \quad - H(c (\al_2-\al_1)S^N+\Zt_{21}^N|W),
\label{eq:uniform 3 SECOND group}
}{\label{eq:uniform 3 SECOND group all}}
where  \eqref{eq:onejmb} follows from the observation in \eqref{eq:set 1} and  \eqref{eq:uniform 3 SECOND group} follows again from the fact that this transformation has unitary Jacobian.
Similarly to \eqref{eq:set minus one}, we have that the set
\ea{
\lcb \{ Y_i-Y_{(2),i}, \  a_i=\al_1 \}, \{ Y_{(2),i}-Y_{(1),i}, \  a_i=\al_2 \}, \{ Y_{(1),i}-Y_i, \  a_i=\al_3 \} \rcb,
}
contains the elements of the vector
\ea{
c (\al_3-\al_1)S^N+ \Zt_{31}^N,
}
for $\Zt_{31}^N$ defined similarly as in \eqref{eq:Z21}.
With this observation, we write
\eas{
& - 3 H(Y^N|W,A^N=a^N)   \nonumber \\
& \tiny{ \leq H \lb \{Y_i, \  a_i=\al_1 \}, \{ Y_{(2),i} \  a_i=\al_2 \}, \{Y_{(1),i} \  a_i=\al_3 \} | W,c (\al_3-\al_1)S^N+\Zt_{31}^N, c (\al_3-\al_1)S^N+\Zt_{21}^N, A^N=a^N \rb  \nonumber} \\
& \quad \quad- H(c (\al_3-\al_1)S^N+\Zt_{31}^N|W,c (\al_2-\al_1)S^N+\Zt_{21}^N) \nonumber \\
& \quad \quad\quad \quad - H(c (\al_2-\al_1)S^N+\Zt_{21}^N|W) \nonumber \\
& \leq H \lb \{Z_i, \  a_i=\al_1 \}, \{ Z_{(2),i} \  a_i=\al_2 \}, \{Z_{(1),i} \  a_i=\al_3 \} | W,S^N,\Zt_{21}^N, \Zt_{31}^N,A^N=a^N \rb
\label{eq: M3 sigle letter -1}\\
& \quad \quad- H(c (\al_3-\al_1)S^N+\Zt_{31}^N|W,c (\al_2-\al_1)S^N+\Zt_{21}^N)
\label{eq: M3 sigle letter 0}\\
& \quad \quad\quad \quad - H(c (\al_2-\al_1)S^N+\Zt_{21}^N|W),
\label{eq: M3 sigle letter}
}{\label{eq: M3 sigle letter total}}
We are now left with the task of evaluating the terms in \eqref{eq: M3 sigle letter -1}, \eqref{eq: M3 sigle letter 0}
and \eqref{eq: M3 sigle letter} in closed-form.
For the term in \eqref{eq: M3 sigle letter}, we have write
%
%
\eas{
 H(c (\al_2-\al_1)S^N +\Zt_{21}^N)
&=-\f N 2 \log 2 \pi e \lb c^2 (\al_2-\al_1)^2 +2   \rb  \nonumber \\
& \leq -\f N 2 \log 2 \pi e \lb c^2(c^2-1) \al_1^2  +2   \rb  \label{eq:a2-a1 0} \\
& \leq -\f N 2 \log 2 \pi e \lb c^2 +1  \rb-\f 1 2,
\label{eq:a2-a1}
}{\label{eq:a2-a1 tot}}
where \eqref{eq:a2-a1 0} follows from the fact $\al_2> c \al_1$ and \eqref{eq:a2-a1} follows from  $\al_1>1/(c-1)$
as prescribed by  \eqref{eq:strong fading intro}.
For the term in \eqref{eq: M3 sigle letter 0}, we have
\eas{
 & -H(c (\al_3-\al_1)S^N+\Zt_{31}^N|W,c (\al_2-\al_1)S^N+\Zt_{21}^N)\\
& =-N H \lb c^2(\al_3-\al_1)(\al_2-\al_1) \lb 1 -\f{c (\al_2-\al_1) }{ c^2 (\al_2-\al_1)^2 +2 } \rb S + \Zt_{13} - \f{c^2 (\al_3-\al_1)(\al_2-\al_1)}{c^2 (\al_2-\al_1)^2 +2 } \Zt_{12}   \rb \nonumber  \\
& \leq - \f N 2 \log 2 \pi e \lb  1+c^2 \f {(\al_3-\al_1)^2}{2+c^2(\al_2-\al_1)^2} \rb   \nonumber  \\
& \leq - \f N 2 \log 2 \pi e \lb  1+c^2 \f {a^2(c-1)^2}{2+c^2(\al_2-(c-1)^{-1})^2} \rb   \nonumber  \\
& \leq - \f N 2 \log 2 \pi e \lb  1+\f 1 2 c^2 \rb,
\label{eq:a3-a1}
}{\label{eq:C2}}
where, in \eqref{eq:a3-a1}, we have used the fact that $\al_3> c \al_2$, $\al_1>1/(1-c)$ and $c>2$ by assumption.
Finally, the term in  \eqref{eq: M3 sigle letter -1} only contains independent noise terms, so that
\ea{
H \lb \{Z_i, \  a_i=\al_1 \}, \{ Z_{(2),i} \  a_i=\al_2 \}, \{Z_{(1),i} \  a_i=\al_3 \} |\Zt_{21}^N, \Zt_{31}^N , A=a^N \rb
& =\f N 2 \log  \lb \f 1 3 \rb.
\label{eq:C3}
}
Combining the bounds in \eqref{eq:a2-a1 tot}, \eqref{eq:C2} and \eqref{eq:C3} we finally obtain the
%
outer bound
\ea{
R^{\rm OUT} \leq \f 1 2 \log (P+c^2(1+\mu_A)  +1) - \f 3 4 \log(c^2 (1+\mu_A)) + \f 3 4 \log(1+\mu_A).
\label{eq:outer before}
}

The final outer bound expression in \eqref{eq:outer strong} is obtained by using Lem. \ref{lem:capacity decreasing c} to tighten the expression
\eqref{eq:outer before}  with the appropriate choice of $c$.
The gap between inner and outer bound is obtained similarly to Th. \ref{thm:approximate capacity discrete mass>1/2}.

\noindent
$\bullet$ {\bf General converse.} The derivation for the case $M>3$ is obtained by extending the derivation for $M=3$ as follows.
First, we define $M-1$ conjugate sequences as
\ea{
\ao_{(k)}^N(a^N) = \lcb a_i=\al_j \implies  a_{(k),i}=\al_{ {\rm mod}(k+j,M)} \rcb,
\label{eq:definition ao}
}
for $k\in[1\ldots M-1]$.
Next, the bounding in \eqref{eq: M3 sigle letter total} can be repeated recursively $M-1$ times: this yields $M-1$ terms of the form
$H(\Delta_i S + \Zt_i | \Delta_1 S +\Zt_1 \ldots \Delta_{i-1} S +\Zt_{i-1})$  for
$\Delta_i=\al_{i+1}-\al_1$, $\Zt_{i1}=Z_i-Z_1$ as in \eqref{eq:Z21}, and
\ea{
H(\Delta_i S + \Zt_{(i+1)1} | \Delta_1 S +\Zt_{21} \ldots \Delta_{i-1} S +\Zt_{i1})
= \f 12 \log \lb 2 \f{ c^2 (\sum_{j=1}^i \Delta_j^2)+2}{c^2 \lb \sum_{j=1}^{i-1} \Delta_j^2 \rb +2} \rb.
\label{eq:recursion delta}
}
%
%
%
The conditions in \eqref{eq:outer strong} guarantee that
\ea{
H(\Delta_i S + \Zt_{(i+1)1} | \Delta_1 S +\Zt_{21} \ldots \Delta_{i-1} S +\Zt_{i1})  \leq -\f 12 \log(2 \pi e c^2)+\f 12,
}
for each $i \in [1 \ldots M-1]$, yielding an outer bound in the spirit \eqref{eq:outer before}
\ea{
R^{\rm OUT} \leq \f 1 2 \log (P+c^2(1+\mu_A)  +1) - \f {M-1}{M} \log(c^2 )
\label{eq:outer before 2}
}
which can be tightened over the parameter $c$  using Lem.  \ref{lem:capacity decreasing c}.
This tightening step  finally yields the outer bound in \eqref{eq:outer strong}.

\newpage
\bibliographystyle{IEEEtran}
\bibliography{steBib1}
\end{document}